\documentclass{ws-ijqi}
\usepackage{bbm}
\begin{document}
\def\set#1{{\cal #1}}\def\sH{\set{H}}
\def\Tr{\hbox{Tr}}
\def\sT{{\scriptscriptstyle T}}
\def\kket#1{|#1\rangle\rangle}
\def\bbra#1{\langle\langle #1|}
\newcommand{\ket}[1]{ |#1  \rangle}
\newcommand{\bra}[1]{ \langle #1|}
\newcommand{\dket}[1]{ | #1  \rangle\!\rangle}
\markboth{Matteo G A Paris}{Unitary local invariance}
\title{UNITARY LOCAL INVARIANCE}
\author{MATTEO G A PARIS\footnote{{\tt matteo.paris@fisica.unimi.it}}}
\address{Dipartimento di Fisica dell'Universit\`a degli Studi di Milano, Italia.}
\date{\today}
\maketitle
\begin{abstract}
We address unitary local (UL) invariance of bipartite pure states.
Given a bipartite state $|\Psi\rangle\rangle=\sum_{ij} \psi_{ij}\: 
|i\rangle_1\otimes |j\rangle_2$ the complete characterization of the 
class of local unitaries $U_1\otimes U_2$ for which 
$U_1\otimes U_2 |\Psi\rangle\rangle=|\Psi\rangle\rangle$ 
is obtained.
The two relevant parameters 
are the rank of the matrix $\Psi$, $[\Psi]_{ij}=\psi_{ij}$, and the number 
of its equal singular values, {\em i.e.} the degeneracy of the eigenvalues 
of the partial traces of $|\Psi\rangle\rangle$. 
\end{abstract}
\keywords{Unitary local invariance; entanglement.}
$ $ \par\noindent
Suppose you are given a bipartite pure state $|\Psi\rangle\rangle\in
{\cal H}_1 \otimes {\cal H}_2$ and asked for which (pairs of)
unitaries the state is locally invariant {\em i.e.} 
\begin{eqnarray}
U_1 \otimes U_2 |\Psi\rangle\rangle = |\Psi\rangle\rangle
\label{uli}\;.
\end{eqnarray}
This kind of invariance is closely related to the so-called 
environment-assisted invariance (envariance) which has been 
recently introduced \cite{zr1,zr2} to understand the origin 
of Born rule. More generally, unitary 
local (UL) invariance naturally arises whenever one investigates 
the possibility of {\em undoing} a local operation performed on a 
subsystem of a multipartite state by acting, yet locally, on another 
subsystem. The somehow related concept of twin observables has been
also investigated in order to account for the invariance that can be 
observed in measurements performed on correlated systems \cite{H1,H2}.
\par
As we will see any state $|\Psi\rangle\rangle$ is UL invariant
for some pairs of unitaries and, as one may expect, UL invariance 
and entanglement properties of $|\Psi\rangle\rangle$ are somehow 
related. However, there are separable UL invariant states and, overall, 
the characterization of the class of unitaries leading to UL invariance 
is not immediate. In this paper a complete characterization of the 
pairs of unitaries for which a given state $|\Psi\rangle\rangle$ is 
UL invariant is achieved in terms of the Schmidt decomposition of 
$|\Psi\rangle\rangle$ {\em i.e.} of the singular value decomposition 
of the matrix $\Psi$. We will show
that the relevant parameter is the number of terms in the Schmidt
decomposition of $|\Psi\rangle\rangle$ and, in particular, the
number of (nonzero) equal Schmidt coefficients. 
\par
The main result of the paper is given by the ULI Theorem, whereas Lemma 1 
and Lemma 2 contains preparatory results. More details on the relationship 
between UL invariance and entanglement are given at the end of the paper.
\par
Let us start by establishing notation. 
Given a bases $\{\ket{i}_1\otimes\ket{j}_2\}$ for the Hilbert 
space $\sH_1\otimes\sH_2$ (with $\sH_1$ and $\sH_2$ generally 
not isomorphic), we can write any vector $\dket{\Psi}\in 
\sH_1 \otimes \sH_2$ as \cite{lop} 
\begin{equation}
\dket{\Psi}= \sum_{i=1}^{d_1} \sum_{j=1}^{d_2} 
\psi_{ij}\:\ket{i}_1\otimes\ket{j}_2\;,
\label{defC}
\end{equation}
where $\psi_{ij}$ are the elements of the matrix $\Psi$. 
The above notation induces a bijection among states 
$\dket{\Psi}$ in $\sH_1\otimes \sH_2$ and Hilbert-Schmidt 
operators 
\begin{eqnarray}
A=\sum_{ij}a_{ij}\: \ket{i}_2{}_1\bra{j}
\label{defOP}\;
\end{eqnarray}
from $\sH_1$ to $\sH_2$. The following relations are an immediate 
consequence of the definitions (\ref{defC}) and (\ref{defOP})
\begin{eqnarray}
A\otimes B\dket{\Psi} =\dket{A\Psi B^\sT} \:, &\quad& 
\langle\langle A\dket{B}=\Tr[A^\dagger B]\:,  \label{defop1} 
\\ \hbox{Tr}_2 \left[\kket{A}\bbra{B}\right] = AB^\dag \:,
&\quad&
\hbox{Tr}_1 \left[\kket{A}\bbra{B}\right] = A^{\sT} B^*
\label{defop2}
\end{eqnarray}
where $A^{\sT, *, \dag}$ denote transpose, conjugate and hermitian 
conjugate respectively of the matrix $A$ (and of the operator $A$ 
with respect to the chosen basis). $\Tr_j[\ldots]$ denotes the partial 
trace over the Hilbert space ${\cal H}_j$ whereas 
$AB^\dag$ and $A^{\sT}B^*$ in Eq. (\ref{defop2}) should be meant as 
operators acting on ${\cal H}_1$ and ${\cal H}_2$ respectively.
\par
Using Eq. (\ref{defop1}) the condition (\ref{uli}) for  
UL invariance  can be rewritten as $|U_1 \Psi U_2^\sT \rangle\rangle = 
|\Psi\rangle\rangle$, thus leading to the matrix relation
\begin{eqnarray}
U_1 \Psi = \Psi U_2^*
\label{uliM}\;.
\end{eqnarray}
\par
The singular value decomposition of $\Psi$ is given by 
$\Psi=S_1^\sT\Sigma S_2$ where $S_1$ and $S_2$ are unitary
matrices of suitable dimension and $\Sigma$ is the diagonal matrix 
$\Sigma=\hbox{Diag}(\sigma_1,\ldots,\sigma_r,0,\ldots)$
where $\sigma_j$ are the singular values of $\Psi$ {\em i.e.}
the square roots of the eigenvalues of $\Psi^\dag \Psi$; $r$ 
is the rank of the matrix $\Psi$.
The singular value decomposition of $\Psi$ corresponds
to the Schmidt decomposition of $|\Psi\rangle\rangle$
\begin{eqnarray}
|\Psi\rangle\rangle &=& |S_1^\sT\Sigma S_2\rangle\rangle
= \sum_{ij} (S_1^\sT\Sigma S_2)_{ij}\: 
|i\rangle_1\otimes|j\rangle_2 \nonumber \\
&=& \sum_{ij} \sum_{kl} S_{1ki} \Sigma_{kl} 
S_{2lj}\: |i\rangle_1\otimes|j\rangle_2
= \sum_k \sigma_k\: |\varphi_k\rangle_1\otimes 
|\theta_k\rangle_2
\label{svd}\;,
\end{eqnarray}
where 
$|\varphi_k\rangle_1 = \sum_i S_{1ki} |i\rangle_1$ and
$|\theta_k\rangle_2 = \sum_l S_{2lj} |j\rangle_2$
are the Schmidt basis in  $\sH_1$ and $\sH_2$ respectively.
\par 
Let us now switch to Schmidt basis, which will be employed 
for most of the paper. Bipartite pure states are thus represented 
by kets $|\Sigma\rangle\rangle$ where $\Sigma$ is a diagonal matrix. 
The UL invariance relation (\ref{uliM}) rewrites as 
follows
\begin{eqnarray}
R_1 \Sigma = \Sigma R_2^*
\label{uliM1}\;,
\end{eqnarray}
where we have denoted  by $R_j$ $j=1,2$ the matrices corresponding 
to the unitary transformations in the new (Schmidt) basis.
\par
The first step in the characterization of unitaries that leave
invariant a given state $|\Psi\rangle\rangle$ is given by the 
following Lemma 
\begin{quote}
{\bf Lemma 1}: 
Let $U_j$, $j=1,2$ be unitaries in ${\cal H}_j$ and 
$|\Psi\rangle\rangle$ a
bipartite state on ${\cal H}_1\otimes{\cal H}_2$. If $U_1\otimes U_2
|\Psi\rangle\rangle = |\Psi\rangle\rangle$ then $[U_j,\varrho_j]=0$ where 
$\varrho_j$ are the partial traces of $|\Psi\rangle\rangle$ {\em i.e.} 
$\varrho_1= \hbox{Tr}_2[|\Psi\rangle\rangle\langle\langle\Psi |] = 
\Psi\Psi^\dag $ and 
$\varrho_2= \hbox{Tr}_1[|\Psi\rangle\rangle\langle\langle\Psi |] = 
\Psi^\sT \Psi^* = (\Psi^\dag\Psi)^\sT$.
\end{quote}
\begin{quote}
{\bf Proof:} 
\begin{eqnarray}
U_1^\dag \varrho_1 U_1 
&=& \hbox{Tr}_2 \left[\left(U_1^\dag \otimes {\mathbbm I}\right)|\Psi\rangle\rangle
\langle\langle\Psi |\left(U_1 \otimes {\mathbbm I}\right)\right] \nonumber \\ 
&=& \hbox{Tr}_2 \left[\left(U_1^\dag \otimes {\mathbbm I}\right)
\left(U_1\otimes U_2\right) |\Psi\rangle\rangle \langle\langle\Psi | 
\left(U_1^\dag \otimes U_2^\dag\right)\left(U_1 
\otimes {\mathbbm I}\right)\right] \nonumber \\ 
&=& \hbox{Tr}_2 \left[\left({\mathbbm I} \otimes U_2\right)|\Psi\rangle\rangle
\langle\langle\Psi | \left( {\mathbbm I}\otimes U_2^\dag \right)\right] \nonumber \\ 
&=& \hbox{Tr}_2 \left[ |\Psi\rangle\rangle
\langle\langle\Psi |\right]  = \varrho_1  
\label{Lemma1}\;.
\end{eqnarray}
\end{quote}
The proof that $U_2^\dag \varrho_2 U_2 = \varrho_2$ and thus that
$[U_2,\varrho_2]=0$ goes along the same lines. As a consequence of 
Lemma 1 $U_j$ and $\varrho_j$ posses a common set of eigenvectors,
which coincides with the Schmidt basis of $|\Psi\rangle\rangle$ 
in each Hilbert space.
\par
From the above Lemma and from Eq. (\ref{uliM1}) we can already draw 
some conclusions about the UL invariance properties of some 
particular class of quantum states. Let us first consider separable 
states. These states correspond to rank-one matrices $\Sigma=\sigma_1 
\oplus {\mathbf 0}$ and thus they are UL invariant under 
transformation $R_1\otimes R_2$ if 
\begin{eqnarray}
R_1 = e^{i\phi} \oplus V_1 \qquad R_2 = e^{-i\phi} \oplus V_2
\label{sepULI}\;,
\end{eqnarray}
where $\phi$ is an arbitrary phase, and $V_j$, $j=1,2$ are arbitrary 
unitaries, each acting on the $(d_j-1)$-dimensional null subspace of 
${\cal H}_j$, corresponding to zero singular values.
More generally, Eq. (\ref{uliM1}) indicates that each matrix $R_j$
should be written as $R_j=W_j\oplus V_j$ where $W_j$ acts on the 
$r$-dimensional subspace of ${\cal H}_j$ corresponding to the support
of $\Sigma$, and $V_j$ on the complementary $(d_j-r)$-dimensional 
null subspace. The rest of the paper is devoted to investigate the
structure of $W_j$.
\par
Let us first consider $|\Psi\rangle\rangle$ being a maximally entangled
state, then $\Psi$ is unitary with $\Psi=S_1^\sT \Sigma S_2$ and 
$\Sigma = \frac1{\sqrt{d}} {\mathbbm I}_d$, $d=\min (d_1, d_2)$. 
Following Lemma 1 and Eq. (\ref{uliM1}) we have 
that $R_1=R_2^*$, {\em i.e.} $|\Sigma\rangle\rangle$
is UL invariant for any transformation of the form $R\otimes R^*$ with $R$ 
arbitrary unitary, {\em i.e.} $|\Psi\rangle\rangle$ is UL invariant for 
transformations $S_1^\sT R S_1^* \otimes S_2^\sT R^* S_2^*$. This 
relation, in turn, expresses isotropy of maximally entangled states 
\cite{isotro}. 
If $|\Sigma\rangle\rangle$ has the form 
of a maximally entangled state immersed in a larger Hilbert space, then the 
same conclusion holds on the support of $\Sigma$. The two above 
statements, together with Lemma 1, can  be summarized as a necessary 
and sufficient condition by the following Lemma:
\begin{quote}
{\bf Lemma 2}: Let $|\Sigma\rangle\rangle$ be a rank $r$ bipartite pure 
state in ${\cal H}_1\otimes{\cal H}_2$  with $\Sigma = \frac{1}{\sqrt{r}}\: 
{\mathbbm I}_{r}$ then 
$U_1 \otimes U_2 |\Sigma\rangle\rangle = |\Sigma\rangle\rangle$ for 
$U_1$, $U_2$ iff $U_1=U_2^*$ on the support of $\Sigma$ {\em i.e.} 
for 
\begin{eqnarray}U_1 = W \oplus V_1 \qquad U_2 = W^* \oplus V_2 
\label{dprimeULI}\;,
\end{eqnarray}
where $W$, $V_1$ and $V_2$ are arbitrary unitaries on the 
corresponding $r$-dimensional, $(d_1-r)$-dimensional and 
$(d_2-r)$-dimensional subspaces. 
\end{quote}
\begin{quote}
{\bf Proof:} 
Sufficiency: if $U_1 = W \oplus V_1$ and $U_2 = W^* \oplus V_2$ 
then Eq. (\ref{uliM1}) is automatically satisfied and 
$|\Sigma \rangle\rangle$, $\Sigma= \frac{1}{\sqrt{r}}\: 
{\mathbbm I}_{r}$, is invariant 
under the action of  $U_1 \otimes U_2$. Necessity: if 
$U_1 \otimes U_2 |\Sigma\rangle\rangle = |\Sigma\rangle\rangle$
then Lemma 1 assures that $U_j$ and $\varrho_j$ posses a common set 
of eigenvectors, which coincides with the Schmidt basis of 
$|\Psi\rangle\rangle$, {\em i.e.} the support of 
$|\Sigma\rangle\rangle$. This fact, together with 
Eq. (\ref{uliM1}), implies that $U_1=U_2^*$ on the support 
of $\Sigma$. The thesis then follows by completing $U_1$ and $U_2$ 
in a unitary way on the remaining sectors of the Hilbert space.
\end{quote}
We are now ready to state the main result of the paper in the 
form of the following Theorem:
\begin{quote}
{\bf Theorem (ULI)}: 
Let $|\Psi\rangle\rangle$ be a bipartite pure 
state in ${\cal H}_1\otimes{\cal H}_2$, with $\Psi$ of rank $r$, 
and let $r_k$ be the number of $k$-tuple of equal singular 
values, {\em e.g.} $r_1$ is the number of distinct singular 
values, $r_2$ the number of pairs and so on; $r_1+ 2\: r_2+\ldots 
k\: r_k+\ldots = r$. Then $|\Psi\rangle\rangle$ is UL invariant  
{\em i.e} $U_1 \otimes U_2 |\Psi\rangle\rangle = 
|\Psi\rangle\rangle$  iff $U_j=S_j^\sT R_j S_j^*$, $j=1,2$ with
$R_j$ given by 
\begin{eqnarray}
R_1 &=& e^{i\phi_1} \oplus \ldots \oplus e^{i\phi_{r_1}} \oplus
D_1 \oplus \ldots \oplus D_{r_2} \nonumber \\ 
&\oplus& T_1 \oplus \ldots \oplus T_{r_3} \oplus \ldots \oplus V_1
\label{u1}\:, \\
R_2&=&e^{-i\phi_1} \oplus \ldots \oplus e^{-i\phi_{r_1}} \oplus
D_1^* \oplus \ldots \oplus D_{r_2}^* \nonumber \\ 
&\oplus& T_1^* \oplus \ldots \oplus T_{r_3}^*
\oplus \ldots  \oplus V_2
\label{u2}\:,
\end{eqnarray}
where $S_j$, $j=1,2$ are the unitaries entering the singular value
decomposition of $\Psi=S_1^\sT \Sigma S_2$,  $D_1,\ldots,D_{r_2}$ are 
arbitrary $2\times 2$ unitary matrices, $T_1,\ldots,T_{r_3}$ arbitrary 
$3\times 3$ unitary matrices and so on. $V_1$ and $V_2$ are 
arbitrary unitaries in the null subspaces of ${\cal H}_j$ 
corresponding to zero singular values.
\end{quote}
\begin{quote}
{\bf Proof:} 
After moving to the Schmidt basis by $\Psi=S_1^\sT \Sigma S_2$, 
and according to the consideration made before Lemma 1 we can always 
write the ULI requirement as in Eq. (\ref{uliM1}), with 
$R_j = W_j \oplus V_j$, where the $W_j$ are of rank $r$. Then, as a 
consequence of Lemma 1 each $W_j$ can be decomposed into blocks 
acting on the eigenspaces of $\varrho_j$. Inside each eigenspace, 
whose dimension corresponds to the degeneracy of the eigenvalues 
of $\varrho_j$, {\em i.e.} to the multiplicity of each singular 
value of $\Psi$, the matrix $\Sigma$ is proportional to the identity 
matrix.  We can therefore apply Lemma 2 thus arriving at $W_1=W_2^*=W$ with 
\begin{eqnarray}
W=e^{i\phi_1} \oplus \ldots \oplus e^{i\phi_{r_1}} \oplus
D_1 \oplus \ldots \oplus D_{r_2} \oplus T_1 \oplus \ldots 
\label{blocks}\;,
\end{eqnarray}
from which expressions (\ref{u1}) and ({\ref{u2}}) and, in turn, 
the thesis immediately follow. 
\end{quote}
\par
In conclusion, unitary local (UL) invariance of bipartite pure 
states has been addressed and the complete characterization of 
the class of local unitaries $U_1\otimes U_2$ for which 
$U_1\otimes U_2 |\Psi\rangle\rangle=|\Psi\rangle\rangle$ 
has been obtained in terms of the singular values of the 
matrix $\Psi$. The explicit expression of the matrices 
$U_1$ and $U_2$ has been derived.
Maximally entangled states are UL invariant under any 
transformation of the form $U\otimes U^*$ with arbitrary 
$U$ whereas separable states are UL invariant for unitaries 
of the form $\left(e^{i\phi} \oplus V_1\right) \otimes 
\left( e^{-i\phi} \oplus V_2\right)$ with $V_1$ and $V_2$ 
acting on the null subspaces of ${\cal H}_1$ and ${\cal H}_2$
respectively. In the general case, the two relevant parameters 
are the rank of $\Psi$ and the number of equal singular values 
of $\Psi$, which determines the structure of the unitaries on 
the support. 
\par
Let us now go back to the connections between unitary local
invariance and entanglement, in order to stress that the two 
concepts are not straightforwardly related each other. As already 
mentioned, separable states admit locally invariant transformations 
which, however, are just overall phase factors, which are not observable 
neither locally nor globally. Consider the following, more instructive, 
examples: a state with $d$ slightly different Schmidt coefficients can be 
taken to be arbitrarily close to a maximally entangled state, so that the von
Neumann entropy of the reduced states is approximately equal to $\log d$.
The pairs of unitaries that leave it invariant are just opposite phase
shifts diagonal in the two Schmidt bases. 
On the other hand, a state with only one Schmidt coefficient close to $1$
and $d-1$ small and equal coefficients has an exceedingly larger set of
locally invariant transformations, while the von Neumann entropy of its
reduced states can be put arbitrarily close to zero.   
Thus, in general, UL invariance is not equivalent to 
entanglement, though for the pure bidimensional case it may 
become equivalent when supplemented by a suitable squeezing 
criterion \cite{san}. Notice also that the structure of the locally 
invariant  transformations can be used to evaluate the dimension of the
sets of separable and maximally entangled states \cite{kk}.
\par
The author is grateful to the anonymous referees for helpful 
suggestions.


\begin{thebibliography}{99}
\bibitem{zr1} W. H. Zurek, Phys. Rev. Lett. {\bf 90}, 120404 (2003).
\bibitem{zr2} W. H. Zurek, preprint LANL ArXiV: quant-ph/0405161.
\bibitem{H1} F. Herbut F, N. Damnjanovic, J. Phys. A {\bf 33}, 6023
(2000).
\bibitem{H2} F. Herbut, J. Phys. A {\bf 35}, 1691 (2002).
\bibitem{lop} G.~M.~D'Ariano, P.~Lo Presti, and M.~F.~Sacchi, 
Phys. Lett. A {\bf 272}, 32 (2000).
\bibitem{isotro} M. Horodecki and P. Horodecki, Phys. Rev. A 59, 
4206 (1999).
\bibitem{san} A. R. Usha Devi, Xiaoguang Wang, B. C. Sanders, 
Q. Inf. Proc. {\bf 2}, 207 (2003).
\bibitem{kk} M. M. Sino\l\c{e}cka, K. \.{Z}yczkowski, M. Ku\'s, Acta. Phys. 
Pol. A {\bf 33}, 2081 (2002).
\end{thebibliography}
\end{document}